\documentclass{ckm}                 

\confname{Workshop on the CKM Unitarity Triangle, IPPP Durham, April
  2003}

\title{Charming penguins in $B \to \pi \pi $ from QCD light-cone sum rules\thanks{Talk given by B. Meli\'c}}

\author{A Khodjamirian\thanks{On leave from Yerevan Physics Institute, Yerevan, Armenia},
Th Mannel, B Meli\'c\thanks{On leave from
Rudjer Bo\v skovi\'c Institute, Zagreb, Croatia}}

\address{Institut f\"ur Theoretische Teilchenphysik, Universit\"at Karlsruhe, D-76128 Karlsruhe, Germany}

\begin{document}

\begin{abstract}
We use QCD light-cone sum rules to examine 
the $B \rightarrow \pi\pi$
hadronic matrix element of the current-current operator
with $c$ quarks in the penguin topology (``charming penguin'') as a potential 
source of the substantial $O(1/m_b)$ effects. 
Our results indicate that charming penguins do not generate 
sizable nonperturbative effects 
at finite $m_b$. 
The same is valid for the penguin contractions of
the current-current operators with light quarks. The dominant 
penguin topology effects are predicted to be $O(\alpha_s)$. Still, the 
nonperturbative effects at finite $m_b$ 
can accumulate to a visible effect that is illustrated by 
calculating the CP-asymmetry in 
the $B^0_d \to \pi^+\pi^-$ decay. 
\end{abstract}

\maketitle

\section{Introduction}

With the first measurements of the direct CP asymmetry in the $B$-meson system 
by the BaBar and Belle collaborations, 
charmless two-body hadronic $B$ decays have become 
particularly interesting for constraining the CKM matrix elements. 
In particular, $B \to \pi\pi$ and $B \to K \pi$ decays can be used to 
extract the angle $\gamma = arg(V_{ub}^*)$. There are several strategies used 
to determine the $\gamma$ angle from such decays, mainly based on 
the isospin and SU(3) relations. Unfortunately, the theoretical accuracy of these relations is 
limited and it has to be 
improved by calculating the SU(3) breaking effects \cite{Khodjaproc}. 
In the  direct calculation of the relevant hadronic matrix elements 
one has also to resort to approximate
methods. Apart from the naive factorization, which assumes the 
vanishing nonfactorizable interactions, there are methods which 
try to investigate the latter \cite{Sandaproc,BBNS,AK}. 
QCD factorization approach \cite{BBNS} shows that 
in  the $m_b \to\infty$ limit, exclusive $B$-decay amplitudes  
can be expressed in terms of the factorizable part  
and calculable $O(\alpha_s)$ nonfactorizable corrections.
For phenomenological applications it is important to investigate
the subleading effects in the decay amplitudes suppressed by inverse powers of $m_b$.
Especially interesting are ``soft'' nonfactorizable effects, involving 
low-virtuality  gluons and quarks, not necessarily 
accompanied by an $\alpha_s$-suppression.
Quantitative estimates of nonfactorizable 
contributions, including the power-suppressed $O(1/m_b)$ contributions can be obtained \cite{AK} 
using the method of QCD light-cone sum rules (LCSR).

Among the most intriguing effects in charmless $B$ decays are
the so called ``charming penguins''. 
The $c$-quark pair emitted in the $b\to c \bar{c} d (s) $ decay 
propagates in the environment of the light spectator cloud and  
annihilates to gluons, the latter being absorbed in the final 
charmless state. In this, so called BSS-mechanism \cite{BSS} 
the intermediate $c \bar{c}$ loop generates 
an imaginary part, contributing to the final-state strong rescattering phase.
In QCD factorization 
approach \cite{BBNS}, charming penguins are typically 
small, being a part of the $O(\alpha_s)$ nonfactorizable correction 
to the $B\to \pi\pi$ amplitude.  
On the other hand, fits of two-body charmless $B$ decays do not exclude
substantial $O(1/m_b)$ nonperturbative effects of the charming-penguin type
\cite{Rome}.
Therefore, in \cite{KMM} we have investigated 
the effects generated by $c$-quark loops in charmless $B$ decays by using LCSR. 

\section{Charming penguins in LCSR}

The decay amplitude for the $\bar{B}^0\to\pi^+\pi^-$ decay is given by the hadronic matrix element 
$\langle \pi^+\pi^- |H_{\rm eff}|\bar{B}^0\rangle $ 
of the effective weak Hamiltonian
\begin{eqnarray}
H_{\rm eff} &=& \frac{G_F}{\sqrt{2}} \sum_{p=u,c}
V_{pb}V^{\ast}_{pd} \Bigg \{ C_1 {\cal O}_1^p + 
C_2 {\cal O}_2^p \nonumber \\
& & \hspace*{-0.2cm} + \sum_{i=3,...10} C_i(\mu) {\cal O}_i + C_{7\gamma}{\cal O}_{7\gamma} 
+ C_{8g} {\cal O}_{8g}  
\Bigg \}\!,
\label{eq:heff}
\end{eqnarray}
where  $ {\cal O}_1^p = (\overline{d}\Gamma_{\mu}p)(\overline{p}\Gamma^{\mu} b) $ 
and $ {\cal O}_2^p = (\overline{p} \Gamma_{\mu} p)(\overline{d} \Gamma^{\mu} b)$  
are the current-current operators  
($p = u,c$ and $\Gamma_\mu=\gamma_\mu(1-\gamma_5)$),
${\cal O}_{3-10}$ are the penguin 
operators, and ${\cal O}_{7\gamma}$ and ${\cal O}_{8g}$ are the electric dipole and chromomagnetic dipole operator, 
respectively. 
Each operator entering Eq.~(\ref{eq:heff}) contributes to the 
$B\to \pi\pi$ decay amplitude 
with a number of different contractions of the quark lines 
(topologies). 
In the discussion we will mainly concentrate on the operator 
$ {\cal O}_1^c$. For convenience we decompose this operator as 
$ {\cal {O}}_1^c = \frac13{\cal {O}}_2^c + 2 \tilde{\cal O}_2^c $,
extracting the color-octet part 
$ \tilde{\cal O}_2^c =  (\overline{c} \Gamma_{\mu} \frac{\lambda^a}{2} c)
(\overline{d} \Gamma^{\mu} \frac{\lambda^a}{2} b) \, .  $

The LCSR expression for the $B\to \pi\pi$ hadronic 
matrix element of ${\cal {O}}_1^c$ is derived 
from the procedure described in detail in 
\cite{AK,KMU}. One starts by introducing the correlation function:
\begin{eqnarray}
F_{\alpha}^{(\tilde{{\cal O}}_2^c)} &=& 
i^2 \int \!d^4 x \,e^{-i(p-q)x}\!\int d^4 y \,e^{i(p-k)y}  \nonumber \\
& & \hspace*{-1cm} \times \langle 0 |
T \{ j_{\alpha 5}^{(\pi)}(y) \tilde{{\cal O}}_2^c(0) j_5^{(B)}(x) \} | \pi^-(q)
\rangle 
\nonumber
\\
& =& (p-k)_{\alpha} F(s_1,s_2,P^2) + ...\,, 
\label{eq:corr0}
\end{eqnarray}
where $ j_{\alpha 5}^{(\pi)} = \overline{u} \gamma_{\alpha} \gamma_5 d
$ and $ j_5^{(B)} = i m_b \overline{b} \gamma_5 d $ 
are  the quark currents
interpolating pion and $B$ meson, respectively. 
Only the color-octet part of the ${\cal O}_1^c$ operator contributes at the leading 
order. 
The contributions of ${\cal O}_2^c$ 
have to be considered within higher-order corrections. 

The LCSR expression  for the hadronic matrix element
$
A^{(\tilde{{\cal O}}_2^c)}( \bar{B}^0_d \to \pi^+ \pi^-) 
\equiv \langle \pi^-(p)\pi^+(-q) |
\tilde{{\cal O}}_2^c |\bar{B}^0_d(p-q) \rangle
$ is given by
\begin{eqnarray}
f_\pi f_B A^{(\tilde{{\cal O}}_2^c)}( \bar{B}^0_d \to \pi^+ \pi^-)
e^{-m_B^2/M_2^2}  
&=&\int\limits_{m_b^2}^{s_0^B} \!ds_2
e^{-s_2/M_2^2}\!
\nonumber \\
& & \hspace*{-6.0cm} \times \Bigg \{
\!\int\limits_0^{s_0^\pi} ds_1 e^{-s_1/M_1^2}\mbox{Im}_{s_2}\mbox{Im}_{s_1} F(s_1,s_2,P^2)
\Bigg\}_{P^2\to m_B^2}\!,
\label{eq:sumrule}
\end{eqnarray}
where $M_{1}$ and $M_2$ are the Borel parameters in the pion and
$B$-meson channels, respectively.
The parameter $s_0^\pi$ $(s_0^B)$ is the effective threshold parameter of the 
perturbative continuum in the pion ($B$-meson) channel. 
In the sum rule (\ref{eq:sumrule}) we take the 
finite $m_b$ corrections into account, 
but neglect numerically very small corrections of order $s_0^\pi/m_B^2$. 
The corresponding sum rule for the
hadronic matrix elements of 
the current-current operator with the light quarks ${\cal O}_{1}^u$,
$
A^{(\tilde{{\cal O}}_2^u)}_P( \bar{B}^0_d \to \pi^+ \pi^-) 
\equiv \langle \pi^-\pi^+|
\tilde{{\cal O}}_2^u |\bar{B}^0_d \rangle
$ is easily obtained from LCSR for the $\tilde{\cal O}_{2}^c$ operator
by putting consistently $m_c\to 0$. 

\begin{figure}
\hbox to\hsize{\hss
\includegraphics[width=0.4\hsize]{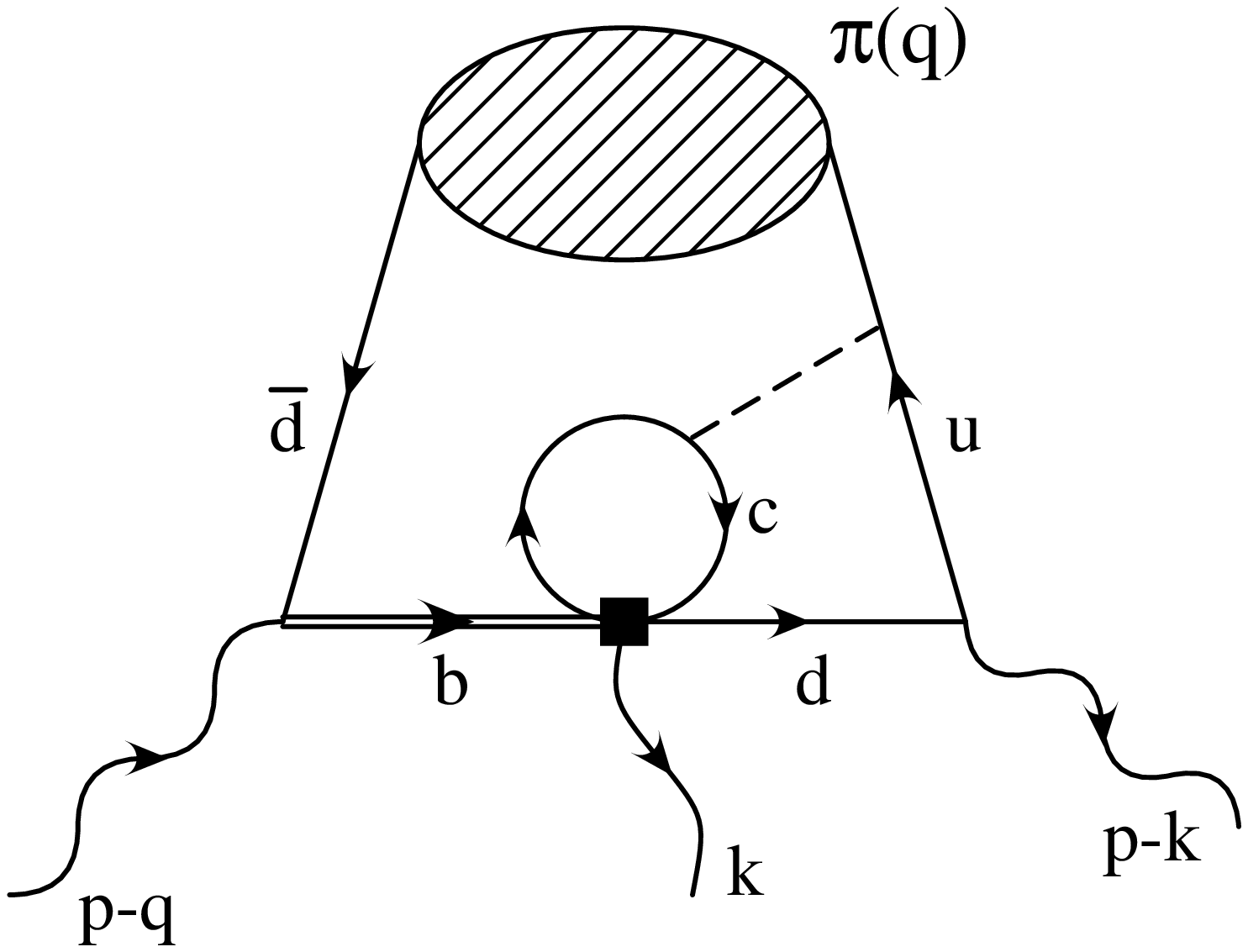}
\hss
\includegraphics[width=0.4\hsize]{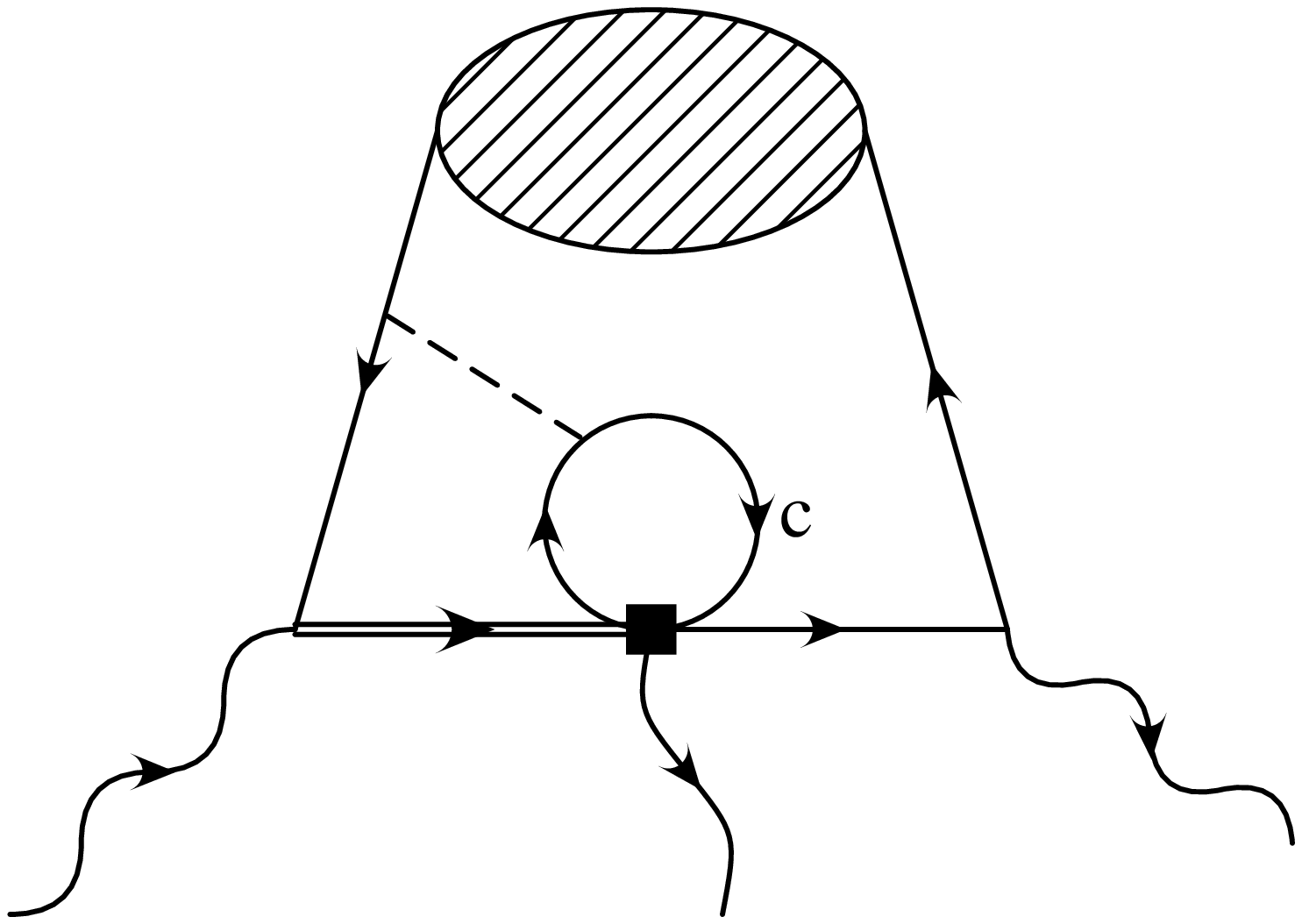}
\hss}
\caption{Diagrams corresponding to
the $O(\alpha_s)$ penguin contractions
in the correlation function (\protect{\ref{eq:corr0}}).
Only the diagrams contributing to the sum rule are  shown.
The square denotes
the four-quark operator ${\tilde{\cal O}}_2^c$.}
\label{fig:hard}
\end{figure}

We calculate $F(s_1,s_2,P^2)$  in (\ref{eq:sumrule}) at large spacelike $s_1,s_2,P^2$
employing the operator product expansion (OPE) near the light-cone.  
The corresponding diagrams of $O(\alpha_s)$  
are shown in Fig.~\ref{fig:hard}. They contain a $c$-quark loop, which
involves a well known function, 
producing the perturbative imaginary part, due to the BSS mechanism \cite{BSS}. 
The remaining diagrams not shown in Fig.~\ref{fig:hard},
with gluons attached to the virtual $b$ and $d$ lines,
do not contribute to the sum rule because their double imaginary
parts vanish inside the duality regions 
$0\!<\!s_1\!<s_0^\pi$, $m_b^2\!<s_2\!<s_0^B$.

%
%

We proceed by investigating the effect 
of the soft (low-virtuality) gluons
coupled to the $c$-quark loop. 
On-shell gluons or light quarks emitted at 
short distances end up in the multiparticle distribution amplitudes (DA's) of the pion. 
These contributions are then of the higher twist and are 
suppressed 
by inverse powers of the heavy mass scale
with respect to the contributions of 2-particle quark-antiquark DA's
of lower twists. Therefore, for our purposes 
it is sufficient to  consider diagrams
with one ``constituent'' gluon i.e., diagrams involving  
quark-antiquark-gluon DA's of the pion. However, for the given 
correlation function (\ref{eq:corr0}), the diagram (Fig. 2a) with one gluon 
vanishes due to the current conservation in the $c$-quark loop.
%
\begin{figure}
\hbox to\hsize{\hss
\includegraphics[width=0.4\hsize]{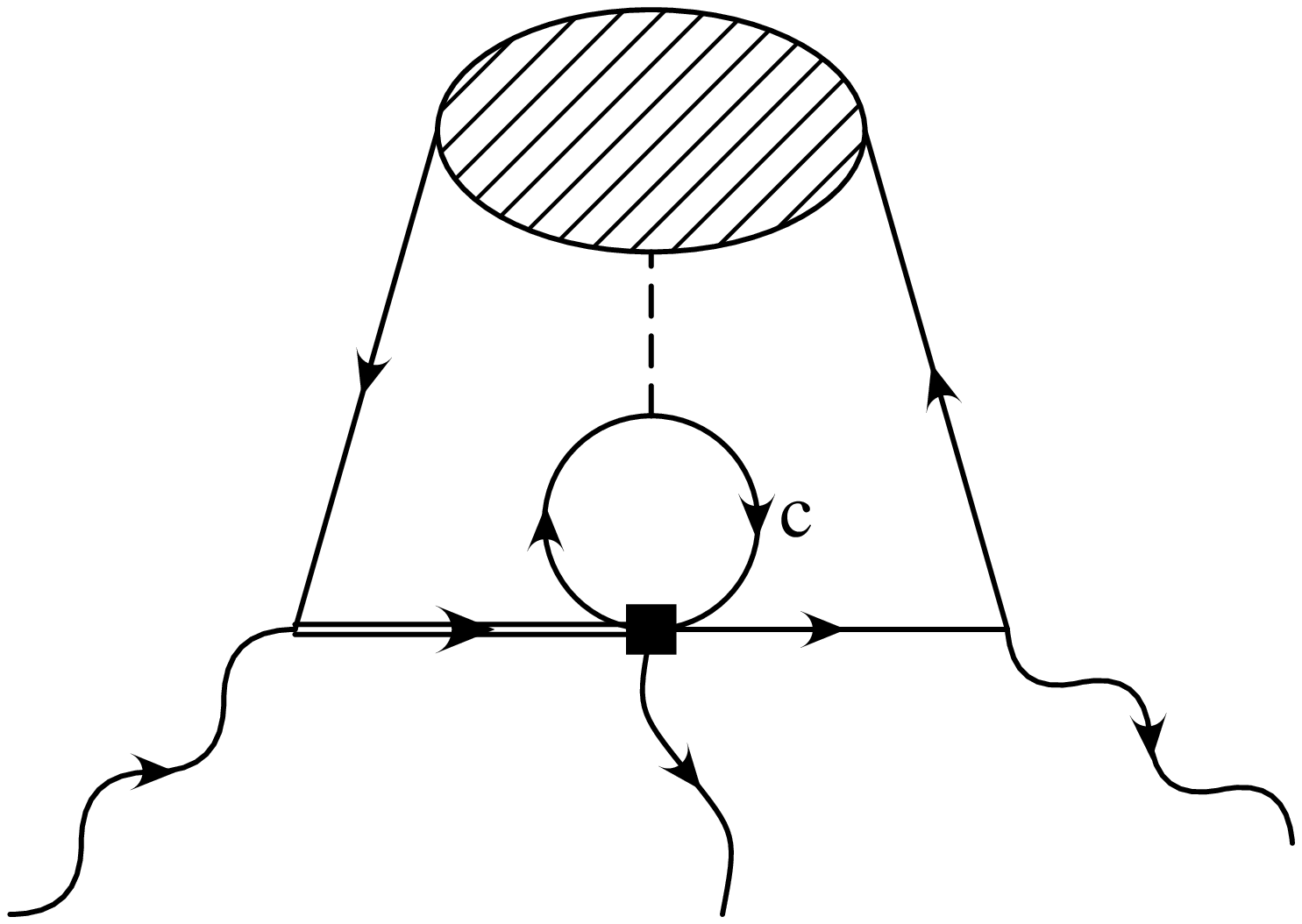}
\hss
\includegraphics[width=0.4\hsize]{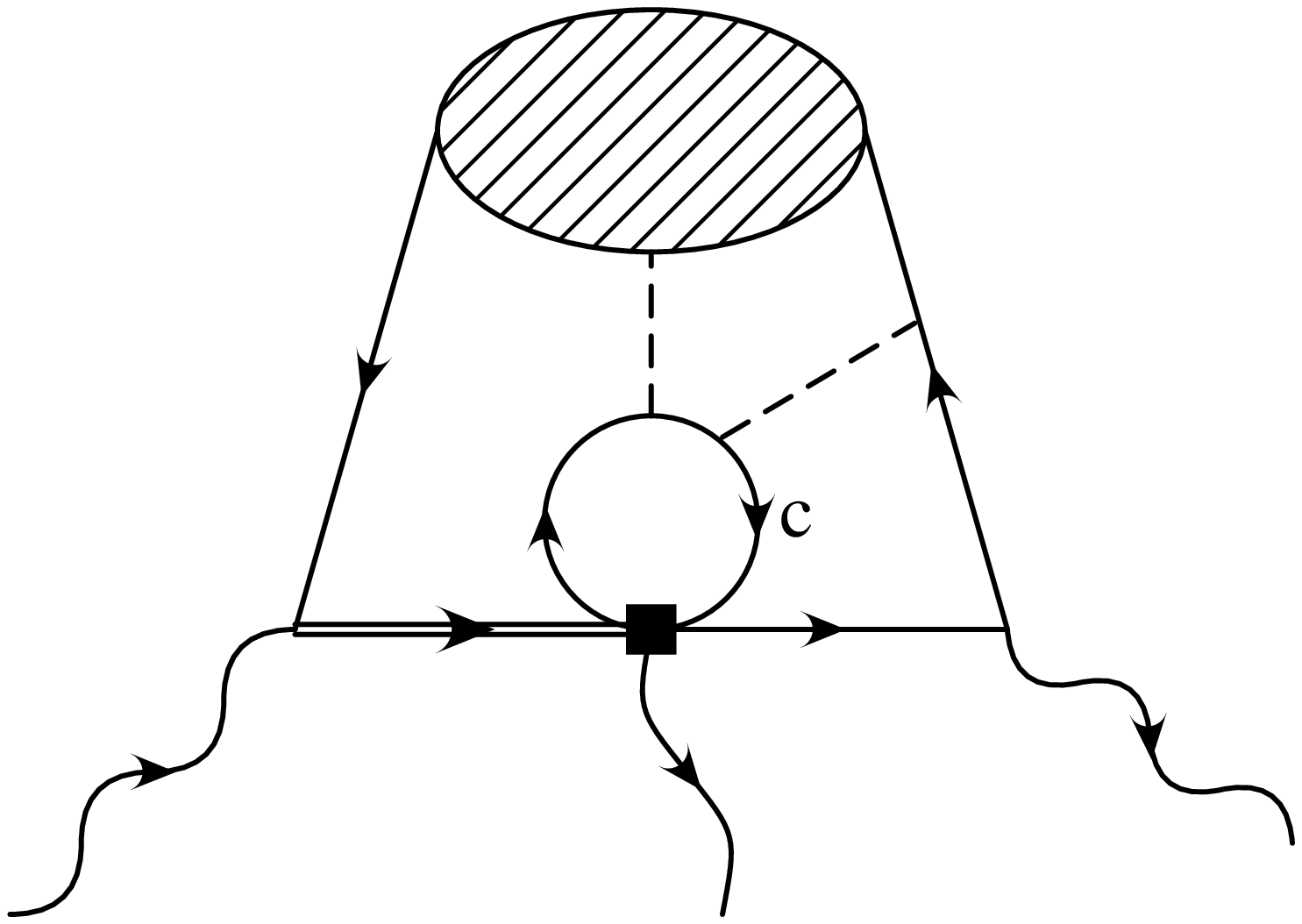}
\hss}
\hspace{2cm} (a)  \hspace{3cm}(b) 
\caption{Diagrams containing the
quark-antiquark-gluon DA's.}
\label{fig:gluon}
\end{figure}
%
%
Nonvanishing terms with the three-particle DA's emerge from the diagrams, 
containing at least one hard gluon in addition to  the on-shell gluon, Fig.2b. 
From the studies of 
the $b\to s\gamma$ matrix elements of ${\cal O}_{1,2}$, 
where similar diagrams with an
on-shell photon and virtual gluon have been calculated \cite{bsgamma}, 
it could be concluded that the contribution of such diagrams  
is both, $\alpha_s$  and $O(1/m_b^2)$ suppressed  
with respect to the diagrams in Fig.~\ref{fig:hard}. 

Furthermore, from the simple 
dimension counting we find that another possible 
contribution, shown in the diagram Fig.3a is 
suppressed by a factor of $O(1/m_b^3) \ln(m_c^2)$. 
%
\begin{figure}
\hbox to\hsize{\hss
\includegraphics[width=0.4\hsize]{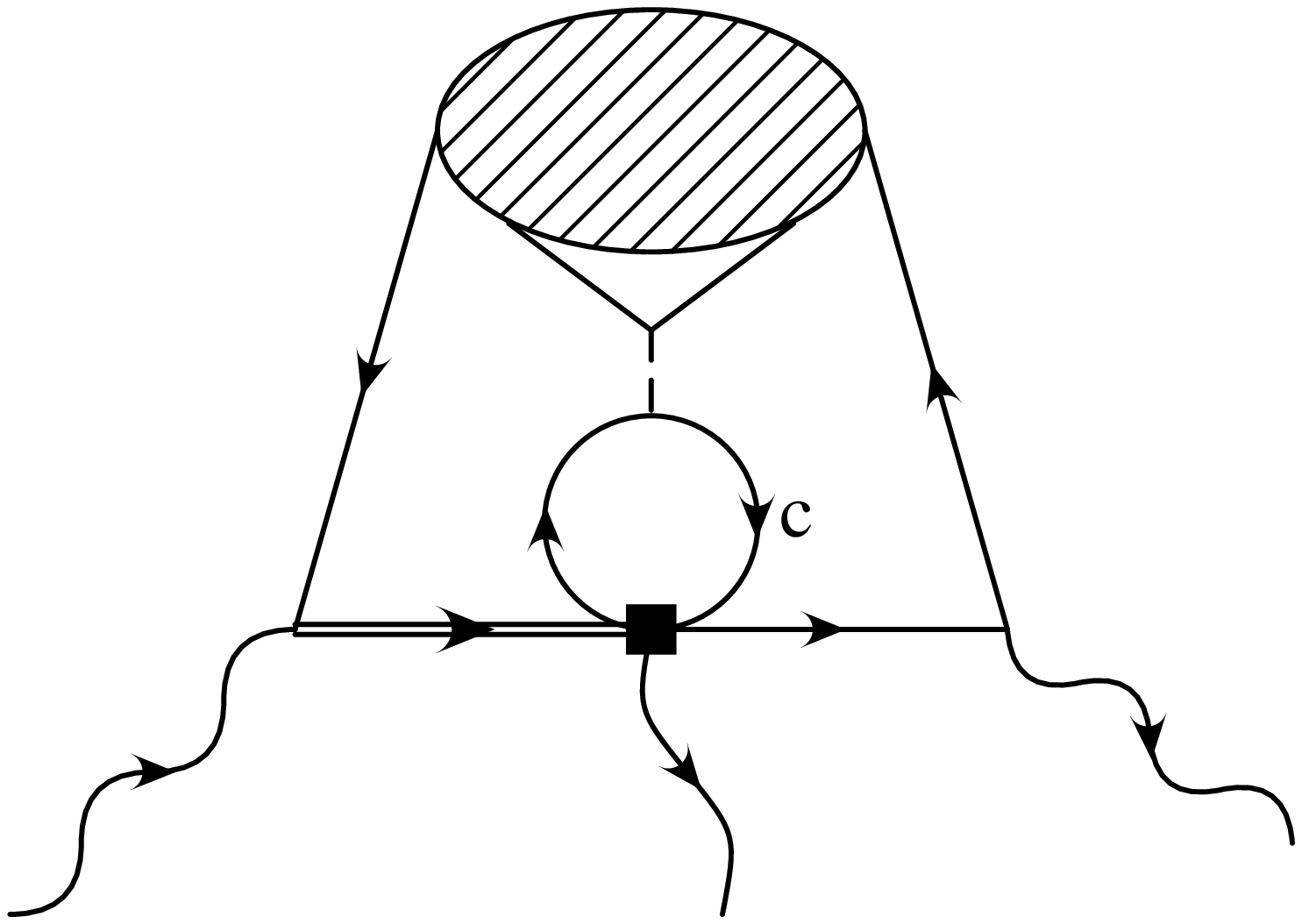}
\hss
\includegraphics[width=0.4\hsize]{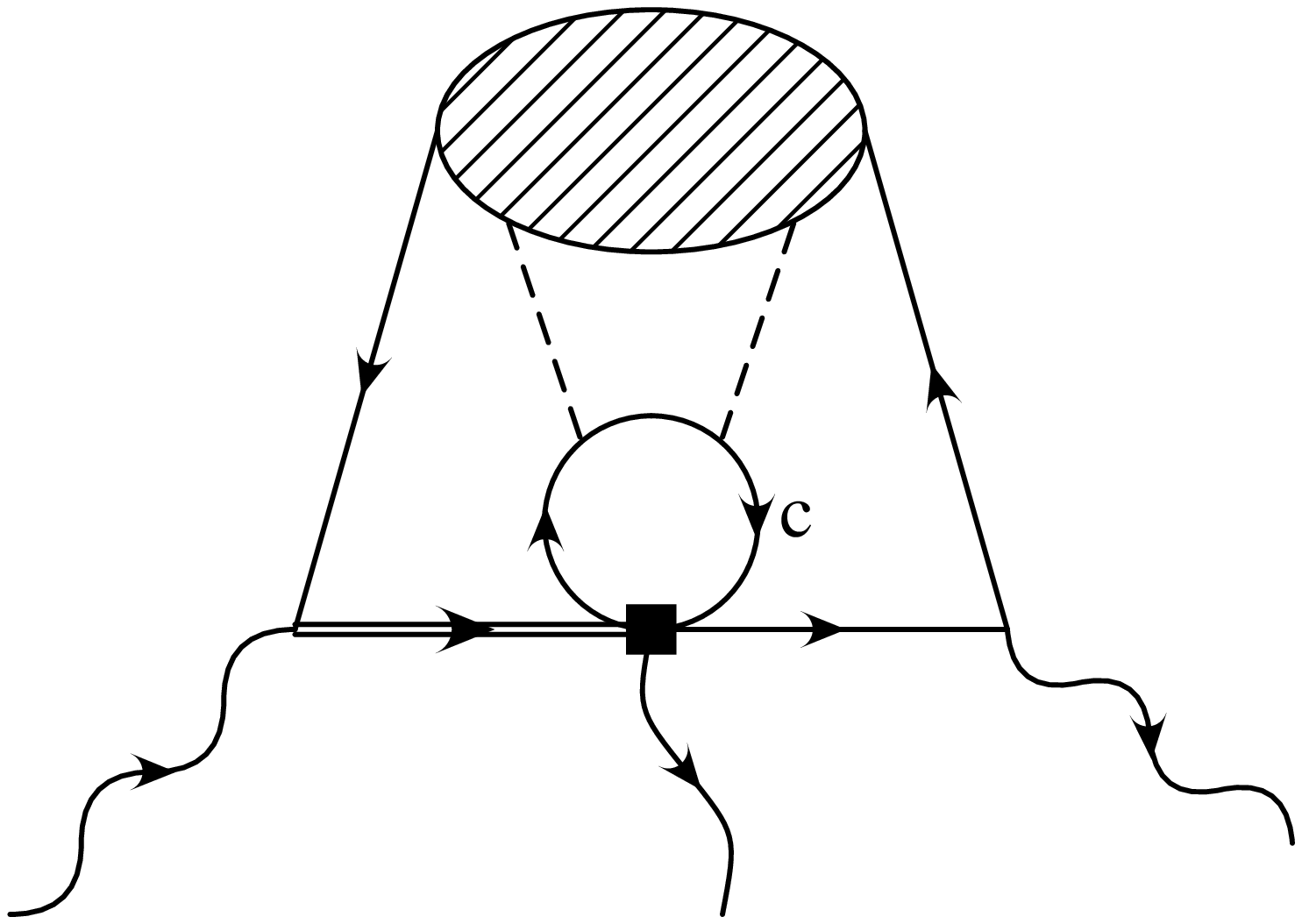}
\hss}
\hspace{1.0cm} (a)  \hspace{2cm}(b)  
\caption{Diagrams corresponding to the multiparticle DA's of the pion.
and arising from the expansion of the $c$-quark loop in the external gluon
field.}
\label{fig:4q}
\end{figure}
%
Also, the diagram with two gluons emitted from the $c$-quark loop (Fig. 3b)
is not included in our calculation because it contains DA's 
with multiplicity larger than three and it is suppressed with 
respect to the diagrams
in Fig.~\ref{fig:hard}  by at least $O(1/(m_c^2m_b^2))$.
The presence of 
$\ln(m_c^2)$ and $m_c^{-2}$ in the contributions of Fig.\ref{fig:4q}a and
Fig.\ref{fig:4q}b respectively, indicates that at $m_c\to 0$  
these terms are divergent and by calculating the contribution 
of ${\cal O}_1^u$ from such diagrams one will have to consider 
propagation of the light-quark pair at long distances, i.e. 
the pion four-quark DA diagram \cite{KMM}.

\begin{figure}
\hbox to\hsize{\hss
\includegraphics[width=0.4\hsize]{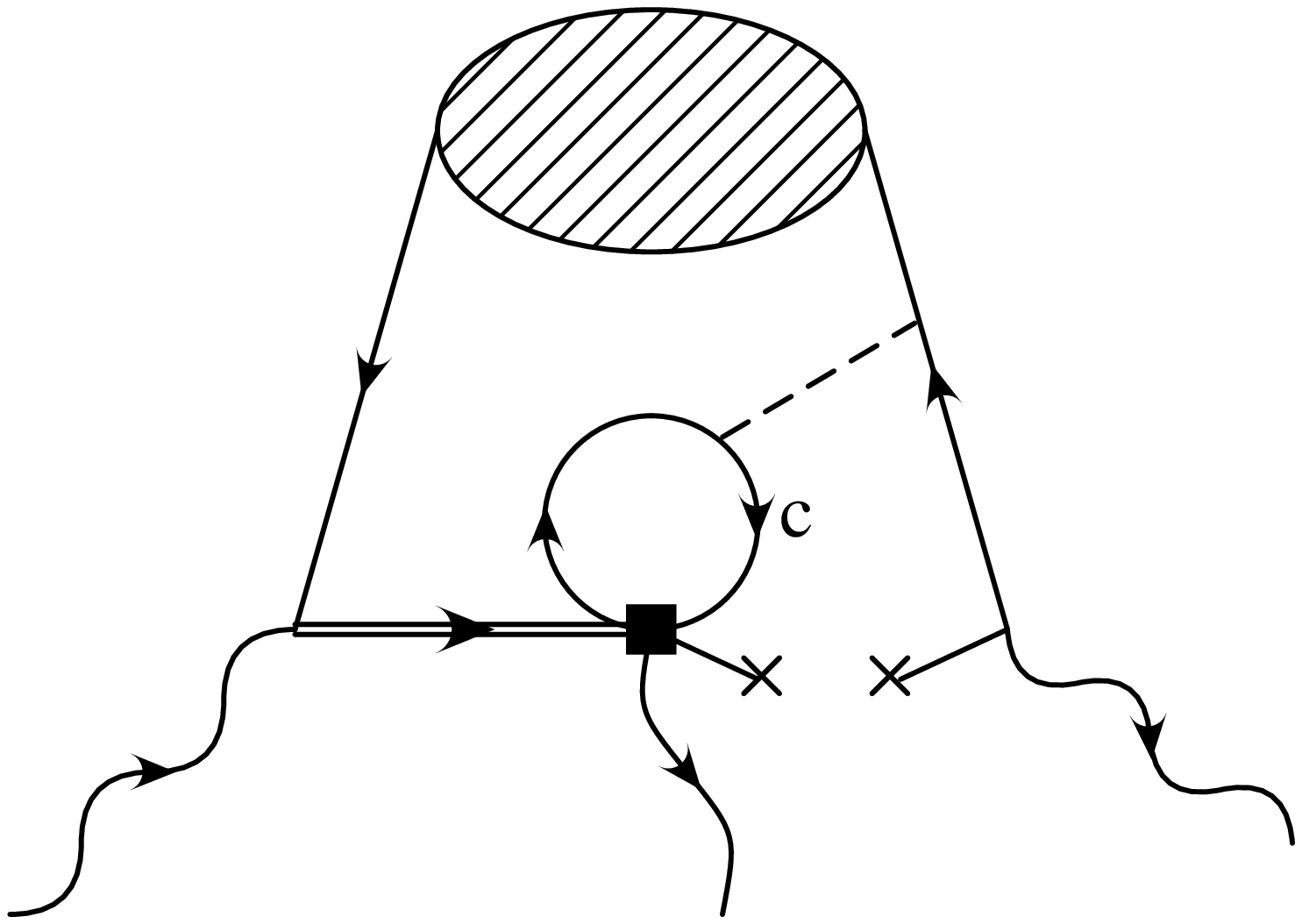}
\hss
\includegraphics[width=0.4\hsize]{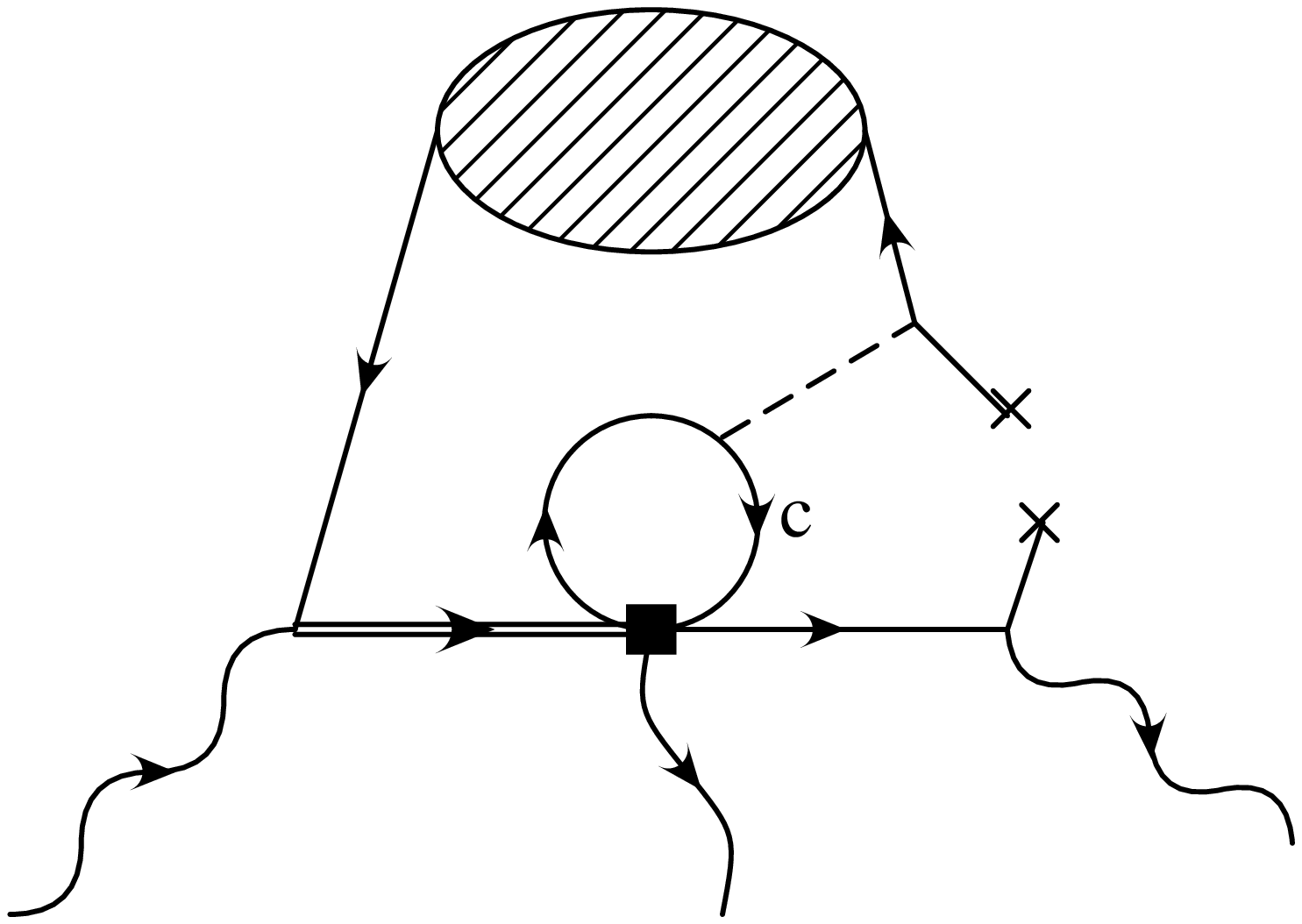}
\hss}
\hspace{2cm} (a)  \hspace{3.1cm}(b) 
\caption{Diagram corresponding to the factorizable 4-quark contribution
to the correlation function.}
\label{fig:cond}
\end{figure}

The contributions of  
four-quark DA's stemming from the matrix elements of the type 
$\langle 0 \mid \bar{u}(x_1)\bar{q}(x_2)q(x_3) d(x_4) |\pi\rangle$ 
($x_i$ on the light-cone) are neglected. But, following \cite{KMU}
we take into account the factorizable parts of the 4-quark
vacuum-pion matrix elements, extracting the configurations 
where one quark-antiquark pair forms the quark vacuum condensate, whereas the other one 
hadronizes into a twist 2 and 3 pion DA. Such 
contributions are enhanced by the  large parameter $\mu_\pi = m_{\pi}^2/(m_u + m_d)$. 
In the approximation adopted in \cite{KMM}, only  
two diagrams shown in Fig.~\ref{fig:cond} contribute to the sum
rule. 
Note that the quark-condensate diagram
in Fig.~\ref{fig:cond}b 
originates from the 4-quark diagram in
Fig.~\ref{fig:4q}a. 
Multiparticle contributions which are factorized in the condensates of higher dimension 
are not taken into account. We also neglect the quark-condensate
contributions of the type $\langle \bar{q}q\rangle \langle 0\mid
\bar{u}(x_1)G_{\mu\nu}^a(x_2) d(x_3)\mid \pi \rangle$ arising from 
the diagram in Fig.~\ref{fig:4q}b after applying the QCD equation of
motion to the derivatives of $G_{\mu\nu}^a$. These terms are suppressed 
at least by $O(1/m_c^2)$ with respect to the diagrams in Fig.~\ref{fig:cond}.
 
To summarize, we do not find significant contributions involving soft
gluons in the OPE of the correlation function (\ref{eq:corr0}). 
The dominant effect which we calculate arises from the $c$-quark loop annihilation into 
hard gluons (Fig.~\ref{fig:hard}). In addition, there is the
quark-condensate contribution (Fig.~\ref{fig:cond}) which we consider 
as a natural upper limit for all neglected contributions of multiparticle DA's. 
%
%
%

\section{Charming penguins and CP asymmetry}

For a numerical estimate of the charming
penguin in $B\to \pi\pi$ decay we calculate 
the ratio of the sum rule (\ref{eq:sumrule})
to the factorizable amplitude 
$A^{({\cal O}_1^u)}_E( \bar{B}^0_d \to \pi^+ \pi^-) 
=im_B^2f_\pi f^+_{B\pi}(0)$:
\begin{eqnarray}
 r^{({\cal O}_1^c)}(\bar{B}^0_d \to \pi^+ \pi^-) &\equiv&
\frac{A^{({\cal O}_1^c)}( \bar{B}^0_d \to \pi^+ \pi^-)}{
A^{(O_1^u)}_E( \bar{B}^0_d \to \pi^+ \pi^-)}
\nonumber \\ 
& \simeq &
\frac{2A^{(\tilde{{\cal O}}_2^c)}( \bar{B}^0_d \to \pi^+ \pi^-)}{
im_B^2f_\pi f^+_{B\pi}(0)^{(LCSR)}}\;.
\label{eq:ratio}
\end{eqnarray}
With the parameters taken from  \cite{KMU,KMM}, and by 
adding linearly the uncertainties caused by the variation of all parameters, 
we get the following for the penguin-loop contractions with $c$ and $u$ quarks in the loop:
\begin{eqnarray}
& & r^{({\cal O}_1^c)}(\bar{B}^0_d\to \pi^+\pi^-)
\nonumber \\ 
& & \hspace*{0.21cm}
= [-(0.29 \div 0.56) -(1.3 \div 1.6)i]\cdot 10^{-2}\,,
\nonumber \\
& & r^{({\cal O}_1^u)}(\bar{B}^0_d\to \pi^+\pi^-)
\nonumber \\
& & \hspace*{0.21cm} = 
[(0.09\div 0.21) -(1.6\pm 2.1)i]\cdot 10^{-2}\,,
\label{eq:numberc}
\end{eqnarray}
  
The charming penguin corrections turn out to be very small,
not larger than the other nonfactorizable corrections. 
However, they appear to produce 
a noticeable effect in the CP asymmetry. Therefore, the CP asymmetry 
appears also to be a good testing ground for the influence of the 
$1/m_b$ corrections in the charming
penguin contributions. 

Penguin contributions influence both the direct and the 
mixing-induced CP violation in $B \to \pi \pi$. 
Here we concentrate on the direct CP asymmetry in $B_d^0 \to \pi^+\pi^-$ decay, which is 
given as 
\begin{eqnarray}
& & a_{\rm CP}^{\rm dir} 
\equiv (1-\left\vert \xi \right\vert^2)/(1+\left\vert \xi \right\vert^2)
\end{eqnarray}
where 
$\xi = e^{-2i(\beta+\gamma)}(1 + R\,  e^{i\gamma})/(1 + R \, e^{-i\gamma})$
and $R \equiv -P/(R_bT)$.
Here $T$ is the contribution to the $B\to \pi\pi$ amplitude proportional to
$V_{ub} V_{ud}^* = |V_{ub} V_{ud}^*| e^{-i\gamma} $. It contains the 
tree amplitude, the penguin-loop contractions of the current-current operators ${\cal O}_{1,2}^u$, and 
also $V_{ub} V_{ud}^*$  proportional penguin operator contractions. 
The remaining contributions, being proportional to $V_{cb} V_{cd}^*$ are contained in $P$. 
The penguin-loop contractions of the current-current operators ${\cal O}_{1,2}^c$ represent the 
main contribution to this part. 
The factor $R_b = |V_{ub}||V_{ud}|/(|V_{cb}||V_{cd}|)$ is the 
ratio of the CKM matrix elements.  

Both $T$ and $P$ amplitudes have strong phases; therefore we have
$T = |T| e^{i\delta_T}$ and $P = |P| e^{i\delta_P}$  and the CP 
asymmetry for $B_d^0 \to \pi^+\pi$ can be written as 
\begin{eqnarray}
& & a_{\rm CP}^{\rm dir} = 
\frac{-2 |R|\,  \sin(\delta_P - \delta_T)\, \sin \gamma}
{1 - 2 |R| \, \cos(\delta_P - \delta_T)\, \cos \gamma + |R|^2}\,.
\label{eq:acp}
\end{eqnarray}
We have to include in $T$ and $P$ all contributions mentioned above. 
The contributions
arising from the penguin contractions of the operators ${\cal O}_1^c$  
and ${\cal O}^u_1$ are taken as in (\ref{eq:numberc}). 
It is easy to extend these results to 
the tree and
penguin contributions
of the penguin operators ${\cal O}_{3-6}$. The LCSR result for the gluonic penguin 
contribution of the dipole operator ${\cal O}_{8g}$ is borrowed from \cite{KMU}.  
The electroweak penguin 
contributions to $B \to \pi \pi$ are color-suppressed and negligible. 

The hard $O(\alpha_s)$ corrections to $T$ and $P$ amplitudes are known in 
$m_b \to \infty $ limit from QCD factorization \cite{BBNS}. We have 
examined the influence of these contributions on the phases $\delta_T$ and $\delta_P$. 
It appeared that they are highly suppressed in comparison to the phases emerging from the 
penguin-loop contributions. Therefore, we neglect $O(\alpha_s)$ corrections in 
(\ref{eq:acp}).

\begin{figure}
\hbox to\hsize{\hss
\includegraphics[width=0.8\hsize]{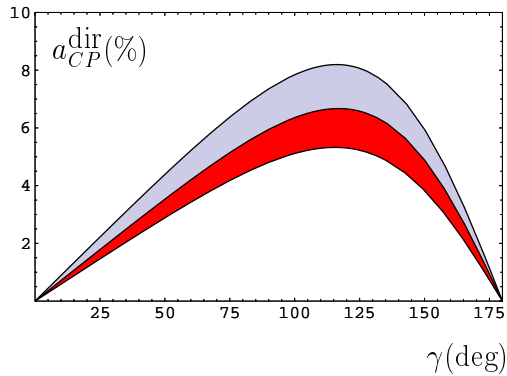}
\hss}
\caption{Direct CP asymmetry in $B_d^0 \to \pi^+\pi^-$ as a function of the CKM
  angle $\gamma$. The upper curve is the result obtained for
$m_b \to \infty$. The dark region is the LCSR result, with all uncertainties from the method included
(uncertainties in the CKM matrix elements are not taken into
account). The light region  shows the deviation from the $m_b \to \infty$
limit result. }
\label{fig:CP}
\end{figure}

In Fig.~\ref{fig:CP} we show $a_{\rm CP}^{\rm dir}$ as a function of 
$\gamma$, calculated by 
using the penguin
contributions estimated from LCSR at finite $m_b$ (dark region) and compare 
the result to the infinite-mass limit that agrees with
the QCD factorization prediction \cite{BBNS} (upper curve). 
Both results are taken at the same scale $\mu_b\sim m_b/2$ as used in LCSR.  
The prediction shown in the figure is not final, because there are missing 
annihilation effects, and the uncertainty in the CKM matrix elements are also 
not taken into account. However, 
the figure nicely illustrates the size of $O(1/m_b)$ corrections  and 
the difference between the result at the finite $m_b$ and
the $m_b\to \infty$ result. 

\section{Conclusion}

The LCSR estimate presented here 
for the hadronic matrix element 
of the current-current operator with penguin topology involving 
$c$ and $u$ quarks  \cite{KMM} 
shows that the 
main contribution to
the sum rule stems from the $O(\alpha_s)$  quark loop annihilating to a
hard gluon, Fig.1. This justifies the generation of the 
strong rescattering phases in $B \to \pi\pi$ by the (perturbative) BSS mechanism.  
The soft-gluon effects, which in the sum rule approach 
correspond to multiparticle pion DA's, are suppressed, at least by
$O(\alpha_s/m_b^2)$. Therefore, we do not find significant nonperturbative 
$O(1/m_b)$ corrections. 
In $m_b\to \infty$ limit our result 
agrees with the QCD factorization prediction for the penguin contractions. 
Since the strong phase is generated perturbatively, 
the CP symmetry in $\bar{B}_d^0\to \pi^+\pi^-$ is expected to be small. 
However, at finite $m_b$ we show that $O(\alpha_s/m_b)$ corrections to 
$a_{\rm CP}^{\rm dir}$ accumulate and can be noticeable. 

\section*{Acknowledgment}
B.M. would like to thank the organizers and the convenors for 
the very interesting 
and pleasant workshop. 
This work is supported by the DFG Forschergruppe 
"Quantenfeldtheorie, Computeralgebra und Monte Carlo 
Simulationen", by the German Ministry for Education and Research (BMBF) 
and by 
the Ministry of Science and Technology 
of the Republic of 
Croatia under the contract 0098002.


\begin{thebibliography}{100}

\bibitem{Khodjaproc}
A. Khodjamirian, Th. Mannel, M. Melcher, in these proceedings. 

\bibitem{Sandaproc}
A. Sanda, in these proceedings. 

\bibitem{BBNS} 
M.~Beneke, G.~Buchalla, M.~Neubert and C.~T.~Sachrajda,
Phys.\ Rev.\ Lett.\  {\bf 83} (1999) 1914;
Nucl.\ Phys.\ B {\bf 606} (2001) 245.


\bibitem{AK}
A.~Khodjamirian,
Nucl.\ Phys.\ B {\bf 605} (2001) 558.



\bibitem{BSS}
M.~Bander, D.~Silverman and A.~Soni,
Phys.\ Rev.\ Lett.\  {\bf 43} (1979) 242.

\bibitem{Rome}
for an overview see 
M.~Ciuchini, E.~Franco, G.~Martinelli, M.~Pierini and L.~Silvestrini,
arXiv:hep-ph/0208048.



\bibitem{KMM}
A. Khodjamirian, Th. Mannel, B. Meli\'c, 
arXiv:hep-ph/0304179.

\bibitem{KMU}
A.~Khodjamirian, T.~Mannel and P.~Urban,
Phys.\ Rev.\ D {\bf 67} (2003) 054027. 

\bibitem{bsgamma}
C.~Greub, T.~Hurth and D.~Wyler,
Phys.\ Rev.\ D {\bf 54} (1996) 3350;






\end{thebibliography}
\end{document}